\documentclass[letterpaper]{article}
\usepackage[utf8]{inputenc}
\usepackage{xcolor}
\usepackage{aaai25} 
\usepackage{times} 
\usepackage{helvet} 
\usepackage{courier} 
\usepackage[hyphens]{url} 
\usepackage{graphicx} 
\usepackage{graphicx} 
\usepackage{natbib} 
\usepackage{caption} 
\newif\ifdraft

\draftfalse 

\newcommand{\added}[1]{%
  \ifdraft
    {\bgroup\color{red}#1\egroup}%
  \else
    #1%
  \fi
}

\title{Principles and Policy Recommendations for Comprehensive Genetic Data Governance}

\author{
\textbf{Vivek Ramanan}\textsuperscript{1},
\textbf{Ria Vinod}\textsuperscript{1},
\textbf{Cole Williams}\textsuperscript{1},
\textbf{Sohini Ramachandran}\textsuperscript{1},
\textbf{Suresh Venkatasubramanian}\textsuperscript{1} \\
{\normalfont\textsuperscript{1}Brown University, Providence, USA} \\
{\normalfont\texttt{\{vivek\_ramanan, ria\_vinod, cole\_williams, sohini\_ramachandran, suresh\}@brown.edu}}
}

\begin{document}

\maketitle

\begin{abstract}

Genetic data collection has become ubiquitous, producing genetic information about health, ancestry, and social traits. However, unregulated use—especially amid evolving scientific understanding—poses serious privacy and discrimination risks. These risks are intensified by advancing AI, particularly multi-modal systems integrating genetic, clinical, behavioral, and environmental data. In this work, we organize the uses of genetic data along four distinct `pillars', and develop a risk assessment framework that identifies key values any governance system must preserve. In doing so, we draw on current privacy scholarship concerning contextual integrity, data relationality, and the Belmont principle. We apply the framework to four real-world case studies and identify critical gaps in existing regulatory frameworks and specific threats to privacy and personal liberties, particularly through genetic discrimination. Finally, we offer three policy recommendations for genetic data governance that safeguard individual rights in today's under-regulated ecosystem of large-scale genetic data collection and usage.

\end{abstract}

\section{Introduction}

The scale of genetic data generation has increased at a staggering rate since 2001 \cite{loos_15_2020, bycroft_uk_2018}. A primary driver of this is decreased DNA sequencing costs \cite{wetterstrand_dna_2023}. For example, there are whole genome sequences combined with medical records from approximately 400,000 US residents in the National Institutes of Health (NIH) All of Us Research Study, with plans to increase sampling to include at least 1 million US residents \cite{nih_all_2024}. Direct-to-consumer (DTC) genetic testing companies have much more data: 23andMe has 14 million customer sequences \cite{23andme_reports_2023} and Ancestry has 25 million \cite{ancestry_company_facts_2025}. The increase in genetic data and improvements in technologies has resulted in the following aspirational ``bold prediction'' for 2030 by the NIH: ``a person's complete genome sequence along with informative annotations can be securely and readily accessible on their smartphone'' \cite{gunter_boldly_2023}. From a scientific perspective, the collection of genetic data \added{and subsequent genetic information} at this scale has been critically valuable to understanding health, ancestry, and driving public health initiatives.

The unregulated collection of genetic data has created new avenues for surveillance, exploitation, and systemic harm. Since 2020, the U.S. Department of Homeland Security has accelerated its collection of genetic data, growing the federal DNA database (CODIS) by more than 1.5 million people, the majority of whom are people of color \cite{glaberson_raiding_2024}. 23andMe suffered a data breach after a hacker accessed customer profiles, primarily targeting individuals of Chinese and Ashkenazi Jewish descent \cite{carballo_23andme_2024}. Such pressing, real-world discrimination and privacy concerns are exacerbated by the sordid history of eugenics and forced sterilization in U.S. states in the 20th century (a model later adopted by Nazi Germany \cite{spiegel_jeremiah_2019}), and more recent practices of genetic determinism \cite{epstein_is_2003}, racial pseudoscience \cite{duello_race_2021, lala_genes_2024, carlson_counter_2022, panofsky_how_2021}, race-motivated violence \cite{carlson_counter_2022}, privacy and consent violations \cite{jillson_dna_2024, sterling_genetic_2011, strand_shedding_2016}, and broad scientific distrust \cite{saulsberry_need_2013, parkman_public_2015, kaye_tension_2012}. These serious concerns must be addressed today given the number of recent proposals to use genetic data and information in public infrastructure—e.g., to assess intelligence proxies, like IQ \cite{r_new_2018}; insurance premiums \cite{karlsson_linner_genetic_2022}; or educational outcomes \cite{harden_genetic_2020}. The only federal-level legislation in effect is the Genetic Information Non-Discrimination Act (GINA), which prohibits using genetic information in employment and health insurance decisions but not in long-term care, disability, or life insurance.

\textbf{Our Work.} In this paper, we make three contributions to construct a robust genetic data governance system. First, we expand upon Wan et al.'s \citeyearpar{wan_sociotechnical_2022} Four Pillars of genetic data collectors and argue that inconsistent and \textit{leaky} regulation across these pillars introduces opportunity for genetic discrimination and privacy violations. Second, we present a risk assessment framework for modern genetic data governance, derived from theoretical analysis and existing U.S. policy. Third, we make policy recommendations to (1) define genetic data and genetic information as a distinct protected category of sensitive data, (2) extend the Genetic Information Nondiscrimination Act (GINA) to include this category, and (3) establish a unified regulatory framework to govern all collection of genetic data and applications of genetic information.


\section{Background: How health and social information is derived from genetic data}

An individual's genome is over 3 billion DNA bases long and derived equally from both genetic parents. Genetic variants—units of DNA that differ between individuals—are scattered throughout the genome and contain information about genetically influenced traits (from physical traits to behavioral traits). To predict traits or disease risk, geneticists analyze an individual's unique set of genetic variants. However, this remains a challenging task: while diseases like cystic fibrosis or Huntington's disease are determined by a single genetic variant, most traits are influenced by many variants across the genome as well as non-genetic factors (complex traits). \added{In this section, we make the connection between genetic data—the raw genomic material—and genetic information, which refers to health, ancestry, and social traits inferred through statistical analysis of that genetic data.} 

\subsection{Genetic data is inherently relational}

The inheritance of DNA—where each parent passes down a random half of their genome to their offspring—results in relatives sharing portions of their DNA, with closer relatives sharing a larger proportion. Shared genetic material means that decisions made about any individual's genome extend to their genetic relatives. At the population level, all individuals can be found in a latent web of interconnectedness. This means that a privacy breach in a small part of this web (that represents those in a genetic dataset, for example) may propagate to larger parts of the web and implicate individuals who were not in the original dataset. Thus, \textit{relationality} applies to genetic data, where value is derived from shared connections among individuals \cite{viljoen_relational_2024, hargreaves_relational_2017}. This relationality complicates discussions of privacy in the context of genetic data \cite{costello_genetic_2022}. 



A powerful example of the relationality of genetic data was highlighted in 2018, when law enforcement leveraged it to identify the Golden State Killer. Authorities used DNA left at a crime scene to identify distant relatives through DNA matching on a public website, GEDMatch. By combining these matches with publicly available genealogical records, they identified the suspect despite the closest identifiable relatives sharing only a great-great-great-great-grandfather \cite{kaiser_we_2018, zabel_killer_2019}. An individual's genome is shared with tens of thousands of other individuals, but at the same time is uniquely theirs: even the genomes of identical twins contain differences \cite{jonsson_differences_2021}. While a single genome in isolation may not always be immediately identifiable, the relational nature of genetic data can enable re-identification. Researchers have argued that genomic data should be treated as ``always identifiable in principle'' due to this inherent interconnectedness \cite{bonomi_privacy_2020, shabani_reidentifiability_2019}.

\subsection{Insight from genetic data is a function of the cohort}



\added{Trait prediction models require a reference genetic dataset, which is analogous to the training dataset in machine learning}. Reference datasets contain genetic data and their ``ground truth'' annotations of trait values, whether it be ancestry, disease status, height, etc. Confounders like environmental exposures and social determinants of health can introduce bias in a genetic study. For example, the average participant in the UK Biobank tends to be more educated and healthier than the average Briton \cite{fry_comparison_2017}. The choice of the \textit{cohort}—that is \textit{who} is in the dataset and how their data is labeled—is thus \added{a critical consideration of trait prediction models and their generalizability, i.e., their out-of-distribution (OOD) performance in individuals outside of the cohort.} Prioritizing ``cohort matching'' in a genetic study ensures that the test sample closely aligns with the reference cohort in key factors such as ancestry, age, geography, and social or environmental influences. This alignment minimizes confounding and enables a more accurate interpretation of strictly genetic effects on traits.

\subsection{Trait prediction models from genetic data}


Inference from genetic data requires some knowledge of a genetic variant's effect on a trait or disease. This can sometimes be tested in animals by controlled experiments that manipulate the animal's genome. However, this approach is infeasible and unethical to pursue in humans, and so researchers instead rely on large genetic datasets to ask ``Do individuals who have the variant \textit{tend} to have the disease (or trait of interest) compared to those who do not have the variant?'' To answer this question, researchers typically conduct genome-wide association studies (GWAS), which identify genetic variants that are associated with a trait and quantify the \textit{effect} of each variant on the trait. The learned effect sizes of variants (also called \textit{weights}) are directly interpretable and can be used for trait prediction. The polygenic score (PGS), considered the gold standard for trait prediction, is calculated as a weighted sum of genetic variants, with each variant scaled by its corresponding GWAS effect size. GWAS/PGS approaches have been widely adopted for trait prediction.





The deployment of AI technologies for genetic inference is nascent but growing: predictive models are increasingly deployed to enable precision medicine therapies \cite{10.1158/1538-7445.AM2025-2773}. Recently, there have been several calls to integrate genetic data with other information—rich clinical, behavioral, and environmental data using foundation models. Critically, biological data—particularly genomic sequences—is large-scale and shaped by noise, sparsity, and biased sampling across the diversity of life. In the domains of vision and language, integrating multi-modal data has shown success in improving predictive performance on sparse and noisy data domains. In the medical domain, Google's Med-Gemini processes electronic health records (EHRs), imaging data (e.g., pathological, dermatological, and X-ray images), and genomic data (in the form of PGS) for trait prediction and other clinical uses \cite{yang_advancing_2024}. They claim Med-Gemini-polygenic—their model for predicting traits with PGS—to be highly performant on OOD data, a highly valued result in AI models for biology. However, it is important to consider that genetic data is often correlated with and confounded by environmental or non-biological variables. Thus, while OOD traits may appear to be genetically correlated with in-distribution (ID) traits (implying shared biological pathways), these correlations—particularly for behavioral phenotypes—are often confounded by non-genetic factors that mimic biological signal, undermining their interpretability as true genetic correlations \cite{border_cross-trait_2022}. Examples of this phenomenon have been demonstrated in places like genetic risk testing for opioid use disorder, where environmental covariates confounded genetic data \cite{pozzi_automated_2023, hatoum_ancestry_2021, davis_utility_2025}.


\subsection*{Blindspots for AI Applications Using Genetic Information}

\added{We believe that the increasing inclusion of genetic data into foundational models—while the interpretability of even simple linear models remains opaque—is of grave concern. GWAS are fraught with confounding: the effect sizes may be statistical artifacts due to correlations between genetic variants and environmental exposures; this is particularly true with behavioral traits. If variant-environmental correlations are not present in another population, the model will generalize poorly. Further, the non-zero effect size of the correlated variant will give the illusion of a genetic or biological basis for the trait, supporting ideas of genetic determinism. Statistical geneticists have developed study designs that control for almost \textit{all} environmental confounding \citep{veller_interpreting_2024, howe_within-sibship_2022, nivard_neither_2022, okbay_polygenic_2022}, but these studies have not been widely adopted.

At the same time, researchers are using these \textit{existing} studies—which suffer from confounding—to train AI models.} We believe that the lack of consideration for this nuance in both training data curation and biological foundation model evaluations can exacerbate pre-existing issues in genetic data modeling approaches. While AI can improve trait prediction or interpretation, it may do so by modeling nonlinear interactions among correlated variables and leveraging gene–environment correlations. Yet, there remains an outsized excitement to develop increasingly complex AI models with genetic data \citep{yang_advancing_2024}. This can lead to compounded uncertainties due to the lack of interpretability and inherent complexity of AI models. \added{We are particularly concerned about the use of multi-modal AI models incorporating genetic data to predict behavior, educational attainment, and other social outcome traits. PGS have already been developed for a variety of these traits, including educational attainment \citep{okbay_polygenic_2022}, intelligence \citep{r_new_2018, selzam_s_comparing_2019}, standardized testing and mathematics scores \citep{harden_genetic_2020, smith-woolley_differences_2018}, substance use disorder \citep{barr_using_2020}, and aggression \citep{kretschmer_polygenic_2023}. Med-Gemini-Polygenic was applied to health traits (i.e., not the aforementioned traits), and leveraged genetically correlated traits to perform well OOD (in this case, OOD refers to the trait, not the individuals). Genetically correlated health traits may reflect shared biology, but for behavioral and social traits, correlations can be induced by non-biological factors \cite{border_cross-trait_2022} and thus an AI model may not be ``learning'' any biological insight. If even simple genetic models struggle with confounding from factors like environmental exposures and cohort biases, complex multi-modal AI models should not be assumed to inherently distinguish true genetic effects from these confounded associations, and may even leverage them, potentially exacerbating existing issues and compounding uncertainties.}

\section{Background: Current Protections}

\begin{figure}[t]
\centering
\includegraphics[width=\columnwidth]{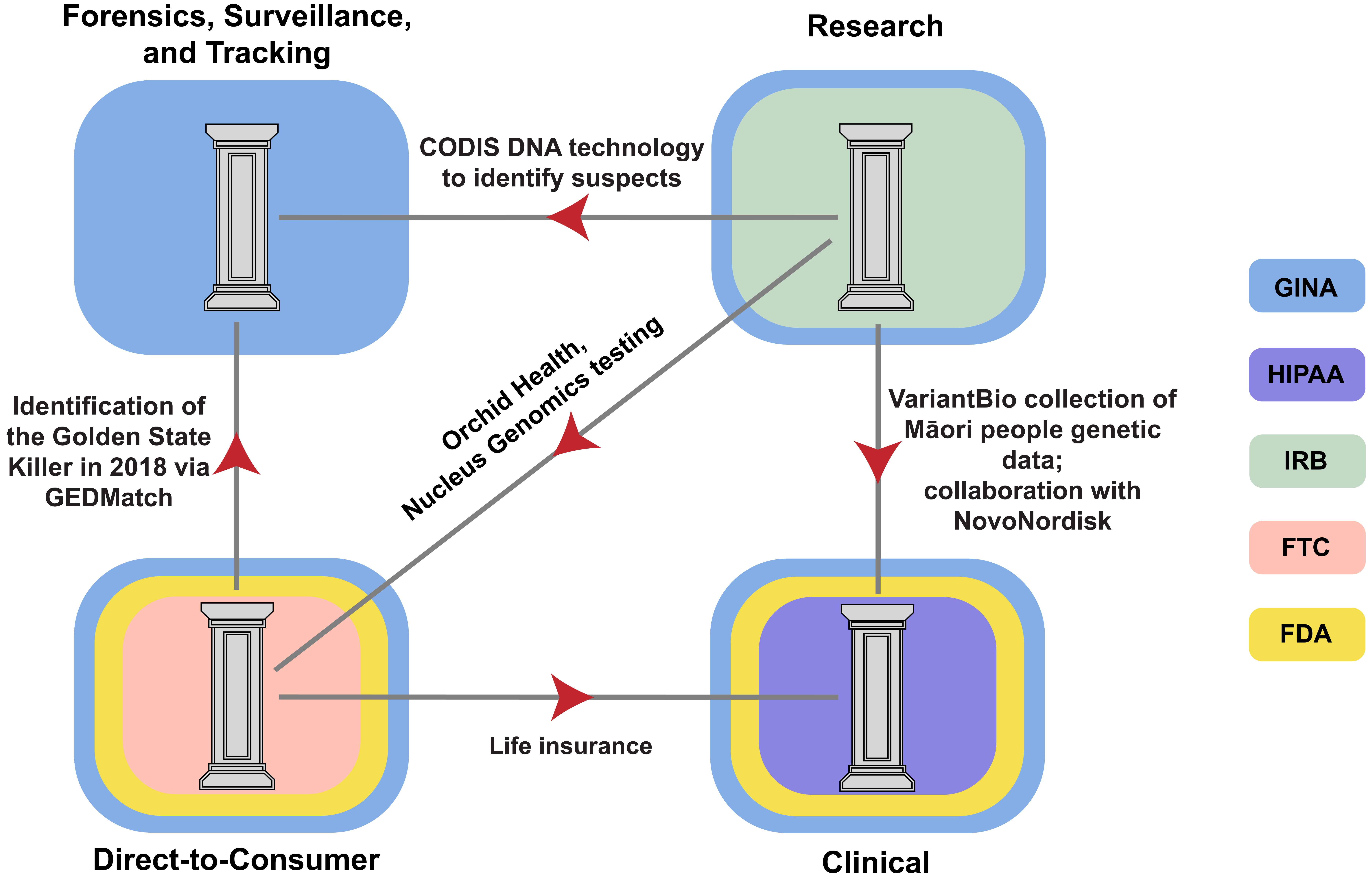}
\caption{The genetic ecosystem consists of the Four Pillars of genetic data collection and regulatory considerations. Data collectors within each pillar operate with different collection strategies and intent to store, use, and transfer the data. However, there is only a patchwork of federal and state-level legislation that governs each Pillar. The resulting regulatory gaps lead to \textit{leaky protections} through which genetic data or genetic information can be used in other pillars with little to no oversight. Real-world examples of leakage are indicated by red arrows. Leakage refers to both (1) genetic data itself and (2) genetic information, i.e., inferences derived from genetic data.}
\label{fig:four_pillars}
\end{figure}

In this section, we outline the foundation of today's genetic data collectors and the regulatory and legislative environments that they are subject to. To better understand the vast landscape over which genetic data is being collected, we build upon the Four Pillars \cite{wan_sociotechnical_2022} of genetic data collectors. We categorize these data collectors based on their motivations for collecting genetic data and the level of consent, control, and ownership afforded to the individuals from whom the data is collected. We define the updated Four Pillars as the following:

\begin{itemize}
\item \textbf{Forensics, Tracking, Surveillance:} Entities focused on monitoring, identifying, and analyzing genetic data to support public health outcomes, security, and law enforcement objectives.
\item \textbf{Direct to Consumer (DTC):} Entities developing products to provide consumers with direct access to genetic insights.
\item \textbf{Clinical:} Entities requiring medical professionals' oversight to conduct tests, often enabling insurance coverage and generating clinically actionable insights.
\item \textbf{Research:} Entities engaging in research and development (R\&D) and basic research across private, public, and academic sectors. This includes public consortia, biobanks, and collaborations with DTC entities.
\end{itemize}

The genetic data ecosystem encompasses not only the entities within the Four Pillars but also the broader network of legislatures, regulators, and individuals involved in or affected by genetic data collection and use. \added{Here, we highlight the regulatory gaps that allow genetic data and genetic information to leak across pillars.} We provide real-world examples of this leakage in Figure \ref{fig:four_pillars}, illustrating the \textit{sociality problem} formalized by Viljoen to contextualize how relational data's social effects are categorized \cite{viljoen_relational_2024}.

\subsection{Federal protections}

\subsubsection{Statute: Genetic Information Non-Discrimination Act (GINA).} GINA was enacted in 2008 and prohibits genetic discrimination by employers or health insurers. Specifically, GINA bars employers from making hiring, firing, or promotion decisions based on genetic information (and information disclosed about relatives) and bans health insurance providers from denying coverage or charging higher premiums based on genetic information, including genetic predisposition for disease \cite{green_gina_2015}. GINA does not apply to life, disability, or long-term care insurance \cite{gunter_boldly_2023} or any other venues outside of employment and health insurance (e.g., education). The US Equal Employment Opportunity Commission (EEOC) regulates workplace discrimination for employees and thus, is responsible for enforcing GINA in the workplace. Before the passing of GINA, multiple cases of civil rights being violated in employment practices using genetic data have been recorded \cite{ajunwa_genetic_2016}. Enforcement for health insurers falls under the purview of several federal agencies, including the Department of Health and Human Services (HHS); Department of Labor; Centers for Medicare and Medicaid Services; and the Department of the Treasury \cite{vermont_legislature_genetic_2019}.\

\subsubsection{Statute: Health Information Portability and Accountability Act (HIPAA).} HIPAA, signed into law in 1996, grants security and privacy protections for patient health information (PHI). HIPAA applies to covered entities that maintain this information, such as healthcare and health insurance companies, but does not apply to many other entities that might retain health information, including search engines, medical information sites, or dating sites \cite{citron_new_2020}. Importantly, HIPAA does not apply to two of the largest DTC genetic testing companies 23andMe and AncestryDNA \cite{sklar_be_2020}. HIPAA does not cover deidentified data: PHI data with unique identifying information (names, address, phone numbers, dates, etc.) removed do not fall under HIPAA. Research consortia, biobanks, and research collaborations with DTC and clinical entities primarily use genetic data that is considered deidentified, and so HIPAA generally does not apply. However, the relational aspect of genetic data has motivated calls to ``consider genomic data as, in principle, always identifiable'' \cite{bonomi_privacy_2020}. The inability to completely strip data of \textit{all} identifiable features introduces critical vulnerabilities for genome owners and exacerbates the risk of data leakage across the Four Pillars.

\subsubsection{Agency: Food and Drug Administration (FDA).} The FDA ``is responsible for protecting the public health by ensuring the safety, efficacy, and security of human and veterinary drugs, biological products, and medical devices'' \cite{commissioner_what_2023}. In the context of DTC genetic testing companies, the FDA only regulates well-defined high risk medical tests with clinical actions, such as 23andMe's health tests for diseases like Parkinson's, breast cancer, and late-onset Alzheimer's. If the FDA approves a specific test, Centers for Medicare and Medicaid Services then choose to approve the test for insurance coverage \cite{daval_authority_2023}. Most genetic tests are not covered by Medicare, particularly DTC tests.

\subsubsection{Agency: Federal Trade Commission (FTC).} The FTC enforces federal laws that protect consumers from ``fraud, deception and unfair business practices'' \cite{federal_trade_commission_enforcement_2024}. The FTC has warned DTC genetic testing companies that the results they return to customers must be backed by ``reliable science'', and also warns companies against making exaggerated claims about the use of AI in their products \cite{jillson_dna_2024}. Additionally, the FTC remains vigilant of deceptive practices regarding privacy policy changes and ``dark patterns'' designed to coerce consumers into consenting to data sharing. The power and broad scope of the FTC—and their stated intentions of cracking down on DTC genetic testing companies \cite{jillson_dna_2024}—make it a major player in the regulatory space, and it will likely grow in importance if the federal government prioritizes genetic data privacy.

\subsubsection{Policy: The Common Rule.} This is a federal policy that protects human subjects in research settings \cite{sciences_federal_2014}. It was first codified in 1981 by the Department of Health, Education, and Welfare (now HHS) but adopted more widely by other federal agencies in 1991. The Common Rule institutionalizes protections for vulnerable participants, protections overseen by Institutional Review Boards (IRBs); IRBs are groups ``formally designated to review and monitor biomedical research involving human subjects'' \cite{research_institutional_2019}. Revisions enacted in 2017 focus on informed consent for participants, but do not cover deidentified data \cite{menikoff_common_2017, protections_ohrp_federal_2009}. IRBs ensure that consistent protection of human subjects are maintained and necessary for research to begin and proceed if involving human participants. The maintenance of the Common Rule and HIPAA, in the case of medical data, is performed by IRBs.

\subsection{State protections}

\begin{figure}[t]
\centering
\includegraphics[width=\columnwidth]{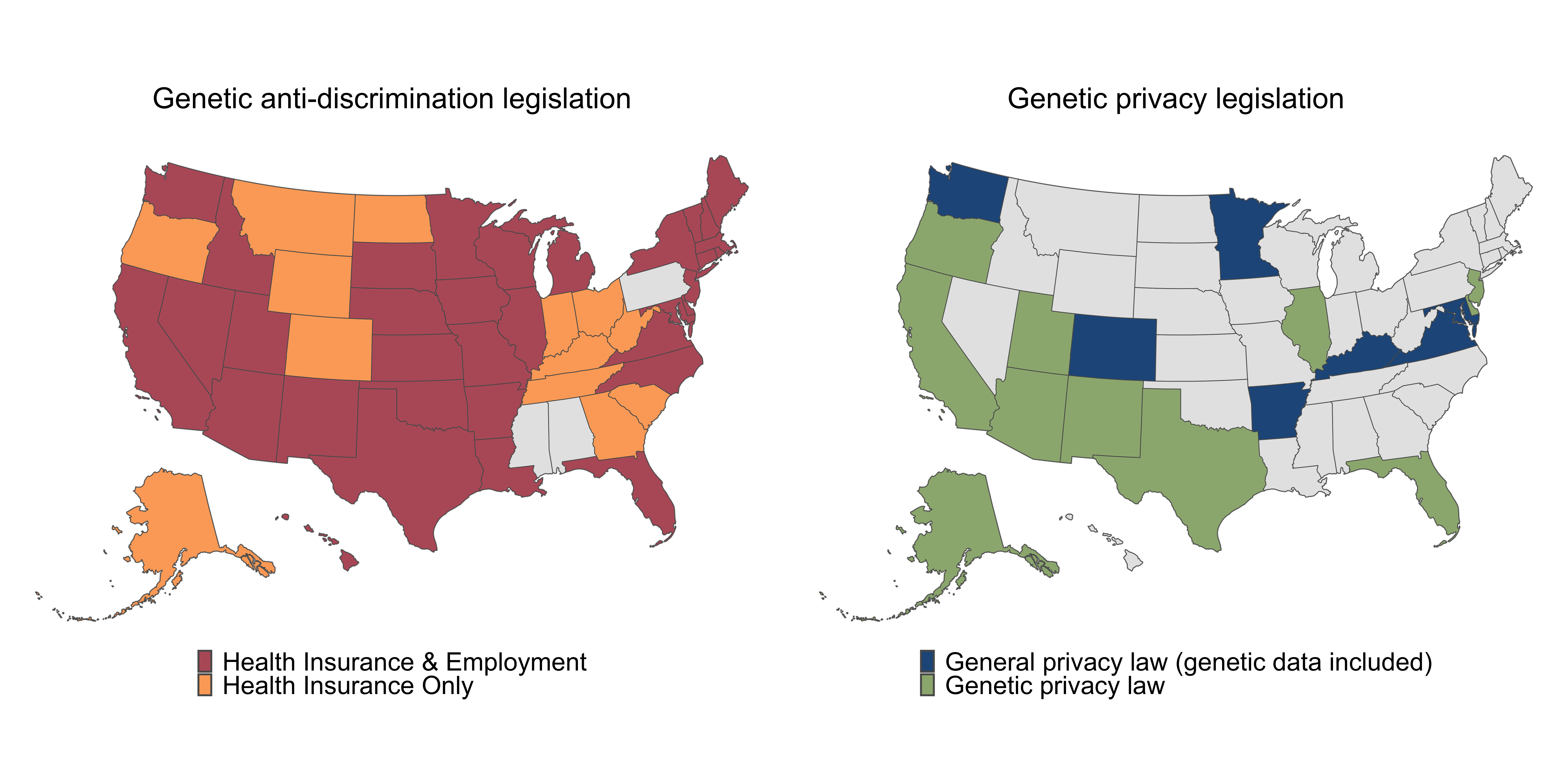}
\caption{State-level genetic discrimination and privacy legislation varies widely beyond federal HIPAA and GINA protections. \textbf{Left}: States with genetic anti-discrimination laws covering both health insurance and employment (red) versus health insurance only (orange). \textbf{Right}: States with general privacy laws including genetic data (blue) and genetic-specific privacy laws (green). States with both are green.}
\label{fig:state_protections}
\end{figure}

State laws offer far more protections than their federal counterparts, but a patchwork of these laws means a patchwork in protections, as shown in Figure \ref{fig:state_protections}. Here, we focus on two categories of laws: genetic anti-discrimination laws and data privacy laws.

\subsubsection{Genetic Anti-Discrimination Laws.} Most states have passed some version of a genetic anti-discrimination law (initially prompted by the passage of GINA) although the sector, scope, and strength of the laws differs between states. Some states offer fewer protections than GINA: Georgia, for example, has a statute outlawing genetic discrimination by health insurers, but not employers. Other states offer more protections than GINA. California's CalGINA \cite{california_senate_bill_559_2011} outlaws genetic discrimination in several sectors: employment, health insurance, housing, and by any state agency or entity receiving state funding (such as emergency medical services). Florida includes life, long-term care, and disability insurers \cite{florida_house_bill_1189_2020}. Many states (including Massachusetts and New Mexico) that include life insurance in their non-discrimination laws exempt companies if genetic data reliably, and based on ``sound actuarial principles'' \cite{massachusetts_genetic_law_2025}, gives information relating to mortality or morbidity (see Section 5.4 for a detailed case study).

\subsubsection{Genetic Data Privacy Laws.} Many states have genetic data privacy laws, either as a standalone law or as part of a broader data privacy law. These laws were generally passed after GINA with the rise of DTC genetic testing companies. Standalone genetic data privacy laws tend to be written specifically to cover DTC genetic testing companies and are often extensions of genetic non-discrimination laws passed around the time of GINA.

Both data privacy and genetic privacy laws have similar provisions \cite{prince_genetic_2021}, including a customer right to request deletion of genetic data and samples and the necessity of consent for data transfers. Some states allow consumers to bring lawsuits for violations of the privacy law (a private right of action). In general though, privacy laws do not apply to deidentified genetic data (as per HIPAA). An important contrast to the US federal and state bills on privacy is the EU General Data Protection Regulation (GDPR). GDPR not only regulates all personal data within the EU but also the transfer of data outside of the EU, data minimization, and compliance. Genetic data is labeled as ``special category data'' in GDPR as it has many identifiers associated with it.

\added{\subsection*{Blindspots for Regulatory Bodies Governing Genetic Data Collection and Usage}}

\added{A critical tension within the current protective framework, predating even advanced AI applications, lies between traditional single-gene tests and newer probabilistic measures of genetic risk, notably polygenic scores (PGS). Legislation like GINA was largely envisioned for single-gene tests that detect specific variants directly linked to conditions (e.g., BRCA1/2 for breast cancer, HTT for Huntington's). These tests offer relatively deterministic insights, which existing laws sought to shield from discriminatory use.

However, PGS operate differently. They don't ``detect'' individual genotypes or mutations in the direct sense that GINA outlines in its definition of ``genetic information'' (an analysis that ``detects genotypes, mutations, or chromosomal changes'' \citep{GINA2008}). Instead, PGS represent a secondary computational analysis, transforming raw genetic data into statistical models of relative risk across populations. This transformation from direct detection to probabilistic, interpretive analysis arguably places PGS outside GINA's explicit protections for employment and health insurance decisions \citep{callier_prince_legal_2024, yanes_future_2024}. This potential loophole is significant because the law focuses on the detection of genetic information rather than the interpretive analyses and risk scores derived from it.

This distinction also manifests in regulatory oversight. The FDA, for instance, regulates well-defined single-gene tests with clear clinical actions, but PGS largely fall outside its purview. While the FTC might address PGS claims under fraud or deceptive practice statutes, this offers a different, and potentially less direct, form of consumer protection than the specific anti-discrimination mandates of GINA or the clinical validation oversight of the FDA.}


\section{Risk Assessment Framework}

Our goal in this section is to develop a risk assessment framework for genetic data governance. We adopt the structure of the aims of the Blueprint for the AI Bill of Rights \cite{park2023ai} and motivate our framework with three central questions: (1) What values (i.e., moral principles and civil liberties) should be preserved? (2) What are the vulnerabilities in the current system that can compromise these values? (3) What are the specific harms that can result from the vulnerabilities?

\subsection{What values should be preserved?}

We seek to identify values that are future-proof and can be encoded in a system of genetic data collection, considering practical ideas of control, consent, and non-discrimination surrounding genetic data. Using Viljoen’s framework of the relational theory of data governance \cite{viljoen_relational_2024}, we identify vertical relationships (between data collectors, i.e. the Four Pillars, and individuals) and horizontal relationships (between genetically connected individuals). This framework must protect personal and civil liberties for both data submitters and their implicated genetic relatives, encompassing both vertical and horizontal relationality. Through this analysis, six key values emerge. The first four values are drawn particularly from Nissenbaum's framework of contextual integrity, highlighting genetic data transfer while preserving personal freedoms \cite{nissenbaum_contextual_2019, nissenbaum_privacy_2004, barth_privacy_2006}.

\begin{enumerate}
\item \textbf{Right to action:} The individual, and the individual only, has the choice to submit (and the freedom to not submit) their genome.
\item \textbf{Ownership of the genome:} The individual owns their genome and therefore controls the usage of their genome, including \textit{who} has access to it and for what reason.
\item \textbf{Right to privacy:} The individual has a right to privacy with regards to their genome and inferences made from their genome.
\item \textbf{Right to knowledge:} The individual has a right to know or \textit{not} know about inferences made from their genome.
\item \textbf{Protecting opportunities for advancement:} Genetic data should not be used to deprive the individual of opportunities in any domain, including (but not limited to) education, access to financial tools, health insurance, housing, social services, and reproductive choices, thus preventing a ``genetic underclass'' \cite{ajunwa_genetic_2016}.
\item \textbf{Benefits of inclusion:} The Belmont principle states ``those who bear the burdens of research (i.e., those who are exposed to the discomforts, inconveniences, and risks) should receive the benefits in equal measure to the burdens'' \cite{belmont_report}. We believe that this applies to individuals and their genetic data.
\end{enumerate}

\subsection{What are the vulnerabilities in the current system that can compromise these values?}

The current genetic data ecosystem has many vulnerabilities that compromise the values listed above. These include:

\begin{enumerate}
\item \textbf{Unsettled science:} The role of genetics in shaping complex traits is poorly understood, and integrating genetic data into multi-modal AI models will likely not improve this understanding. As a consequence, the deterministic nature of genetics may be overstated, particularly for behavioral and cognitive traits.
\item \textbf{Rapid evolution of genetic data/methods:} As genetic data collection, sequencing, and methodologies quickly evolve, legal protections fall behind. For example, education discrimination is not included in GINA, which was passed five years before the first large GWAS on educational attainment \cite{rietveld_gwas_2013}.
\item \textbf{Guilt by association:} DNA databases used for criminal investigations can impact not only the individuals whose DNA is stored, but also their biological relatives, as genetic information is shared among family members.
\item \textbf{Geographical legislative patchwork:} Genetic anti-discrimination and privacy legislation widely differs between states. Legal policy in each state for data removal, third-party sharing, and many other aspects of genetic data are unclear. Individuals who move between states or share data with entities in other states can be affected by this dependency on state specific policy. Aside from patchwork policy, many state policies do not sufficiently protect genetic data specifically.
\end{enumerate}

\subsection{What are the harms that arise as a result?}

In the final section of our risk assessment framework, we outline the consequences that could arise from violations of our key values through the above vulnerabilities. These downstream effects, which we refer to as harms, can not only affect an individual and their immediate genetic relatives and but can also have a broader effect on communities and the population at large.

\begin{enumerate}
\item \textbf{Leakage to the family:} Any conclusion about an individual's genetic trait may be linked back to their relatives, even if no action is taken by their relatives in either (1) submitting data or (2) opting in to receive information. Any derived secondhand knowledge can immediately impact insurance (health, life, etc.) and medical treatment plans, as well as compromise identity.
\item \textbf{Loss of anonymity:} It is straightforward to identify some traits about an individual from their genetic data, and subsequently from their genetic relatives. Common examples include: race, gender, ethnicity, and markers for certain diseases. As the refinement of reference datasets continues and data collection grows, leaky genetic data interfaces can thus compromise information about an individual, their genetic relatives, and unborn children that the individual (and implicated relatives) may wish to maintain as private.
\item \textbf{Loss of Data Control:} Private institutions offering genetic tests control the entire lifecycle of individuals' data from collection and storage to monetization. This control creates significant transparency issues, as it is unclear how data is monetized, which parties benefit, or what agreements include an individual's data. Users may opt out of participation, but unregulated data flows ensure that their genetic footprint persists indefinitely. This lack of transparency is particularly concerning in cases of bankruptcy or acquisition. For example, in 2024, Tempus AI acquired Ambry Genetics, a company that sequences genetic data for 400,000 patients annually \cite{tempus2025}. Tempus stated its intent to ``leverage this data and augment Tempus' current data offering,'' raising questions about how patient data is repurposed and monetized in such deals. Ultimately, an individual's decision to submit genetic data for testing often entails relinquishing full ownership of their genetic information. Without robust safeguards, privacy risks emerge from opaque data practices and leaky interfaces, leaving individuals vulnerable to misuse or unauthorized access.
\item \textbf{Misinformed Actions:} The interpretation of genetic test results and the subsequent actions taken by individuals are profoundly personal, often influencing lifestyle changes, significant financial decisions, and critical medical choices. For instance, individuals with BRCA1 gene mutations indicating an elevated breast cancer risk may opt for preventive interventions. These decisions are particularly sensitive as the interpretation of genetic data evolves alongside advancements in scientific methods. However, the relational nature of genetic data complicates individual autonomy and access to information. It is possible for someone to receive conclusions about themselves indirectly through the test results of genetic relatives. This dynamic, combined with the widespread availability of unregulated genetic tests and the evolving nature of genetic science, increases the potential for harm if individuals act on information that is incomplete, inaccurate, or probabilistic rather than deterministic.
\item \textbf{Financial impact:} Individuals may face financial repercussions if insurance or legal policies fail to provide adequate protection or if coverage is denied. Beyond medical procedures, financial impacts can also arise on a case-by-case basis, such as in situations where genetic data is held for ransom, legal defense is required, or individuals are forced to purchase direct-to-consumer tests (e.g., in order to purchase a life insurance policy).
\end{enumerate}

\section{Case Studies}

We apply our risk assessment framework to four case studies which highlight regulatory gaps across the Four Pillars. These case studies, ranging from past events to speculative scenarios, all involve privacy loss or potential discrimination. We explore case studies that illustrate genetic data leakage in both, a pillar-to-pillar and one-to-many pillar settings.

\subsection{Genetics and Education}

Polygenic scores (PGS) are popular in the social/behavioral sciences for predicting social outcome traits, such as educational attainment (EA; number of schooling years completed by an adult). We will use EA as the example throughout this case study, but note that there is interest in predicting standardized testing scores, mathematics performance, and other traits with substantial environmental influences.

A common metric for assessing PGS accuracy is the percentage of trait variance it explains: higher percentages indicate better predictive performance. A recent 23andMe EA PGS, based on data from over 3 million customers of European descent, explains 12-16\% of the variance in educational attainment (EA) \cite{okbay_polygenic_2022}. This effectively means that 50-70\% of individuals with PGS scores in the top 10\% for EA are predicted to graduate college. However, the PGS accuracy significantly decreases when applied to African American customers. This is an example of the commonly observed ``portability problem'' \cite{martin_human_2017}, where a PGS derived from GWAS in one population predicts poorly in another due to confounding.

There have been several calls to use EA PGS to inform education policy. Harden et al. propose the use of math-performance PGS to identify ``leaks'' in the education system: for example, by identifying high math PGS students who perform poorly, they claim educators could pinpoint why and how students are failing to reach their potential \cite{harden_genetic_2020}. Plomin \& von Stumm take it further: they use the term ``precision education'' (akin to precision medicine) to propose a tailor-made, individualized education that is genetics-informed \cite{r_new_2018}. Statements like this, combined with statements such as ``students with higher polygenic scores for years of education have, on average, higher cognitive ability, better grades and come from families with higher SES [socioeconomic status]'' \cite{smith-woolley_differences_2018} are cause for concern because they invoke a sense of genetic determinism. However, other predictors (parents' educational status, socioeconomic status) explain similar amounts of variance in EA \cite{morris_can_2020, mostafavi_variable_2020} and—unlike DNA—are mutable through social policy changes.

Several of our values would be violated if children were required to submit their DNA (Right to Action) or educational opportunities were denied to children based on their genetic potential (Opportunities for Advancement). Through the vulnerabilities of unsettled science and the rapid evolution of genetic methods, leakage to the family can occur and affect not only children, but their families and future. Crucially, these profound challenges exist even before considering the added complexities and potential for misuse introduced by integrating such scores with advanced AI systems.

\subsection{DTC Genetic Data Brokerage and Leakage}

\added{The fragility of genetic data governance is starkly illustrated by direct-to-consumer (DTC) company 23andMe. Beyond FDA-approved genetic tests (e.g., BRCA1/2 variants), carrier screenings, health risk assessments, and genealogy services, the company also collects extensive non-genetic survey data and conducts its own GWAS.

In 2023, a credential stuffing attack compromised 14,000+ user accounts. Exploiting 23andMe’s relative linkage feature hackers accessed data for an additional 5.5M users \citep{23andme_addressing_2023} in a powerful example of genetic data's relationality. The hackers also reportedly targeted Chinese and Ashkenazi Jewish profiles \citep{carballo_23andme_2024}. A critical factor enabling this breach was the lack of consensus around security responsibility, which resulted in the disproportionate and harmful targeting of underrepresented groups. The breach highlighted how an individual's genetic privacy is hostage to others' security practices, aptly summarized by a Reddit user: ``Your genetic data is only as secure as your relatives' passwords.''

The security and integrity of consumer genetic data was once again compromised when 23andMe filed for bankruptcy and announced its database and assets were for sale \cite{23andme_update_2025}. This prompted widespread calls for data deletion \citep{eddy_23andme_2025} and a federal probe into the sale's privacy and national security risks \citep{house_committee_23andme_2025}. The situation culminated on May 19th, 2025, with the announcement that pharmaceutical giant Regeneron had acquired 23andMe's assets for \$256M \citep{regeneron_23andme_2025}. Regeneron thus gained access to existing genetic data, extensive non-genetic information, model weights, and stored biological samples, which allow for future whole-genome sequencing. Regeneron, in compliance with the terms of the sale \citep{23andme_privacy_2025}, has pledged to uphold 23andMe's terms and conditions (T\&Cs), which include the right to delete data \citep{regeneron_23andme_2025}.

The 23andMe saga reveals critical vulnerabilities and rights violations within our risk assessment framework. The data breach infringed on Right to Privacy, while the asset sale to Regeneron highlighted contested Ownership of the Genome, as users' deletion rights via T\&Cs contrast with the de facto transfer of dataset ownership. This underscores the dangers of a fragmented regulatory landscape: Californians have protections during third-party sales, but those in states like Pennsylvania lack equivalent federal or state safeguards for DTC genetic data sales. Finally, Regeneron's acquisition of biological samples and non-genetic data, especially for use in new AI models, exposes vulnerabilities from rapidly evolving technology and unsettled science, enabling applications far beyond users' original consent or understanding.}

\subsection{Genetic Data Collection of Detained Noncitizens}

A recent Georgetown University report from the Center of Privacy and Technology (CPT) examines the U.S. federal government's practice of collecting DNA from detained migrants \cite{glaberson_raiding_2024}. This practice, which began with the 2005 DNA Fingerprint Act and expanded significantly in 2020—a mandate that the Department of Homeland Security must collect DNA from all detainees, even briefly detained—has led to a dramatic increase in detainee representation in CODIS (the federal DNA database), rising from 0.21\% in 2019 to 9.21\% in 2023. By 2020, approximately 25,000 noncitizens were added to the database under the ``detainee'' classification.

The CPT report highlights several concerning aspects of this practice, particularly its disproportionate impact on migrants of color and issues of consent. In our Risk Assessment Framework, this immediately violates the Right to Action and Right to Knowledge. Many migrants undergo DNA collection without understanding its implications, sometimes believing it to be a COVID-19 test or submitting under threat of criminal prosecution. The report argues that this practice violates the Fourth Amendment in collecting DNA from detainees without probable cause. Additionally, even though CODIS was designed to be privacy-protective by collecting only 20 markers—a tiny slice of the genome believed to be medically neutral—this limited data can still identify relatives through partial matching and, as an analysis 25 years later would show, are slightly informative of disease risk \cite{banuelos_associations_2022}. The expansive collection of genetic data has far-reaching implications through the guilt-by-association vulnerability. Since CODIS records are difficult to expunge, they can restrict Opportunities for Advancement not only for detainees but also for their relatives and descendants.

\subsection{Underwriting Life Insurance with AI and Genetic Data}

Life insurance underwriting is the process whereby an insurance company uses personal and health information to assess the risk of insuring an applicant. The relationship between applicants and insurance companies is already fraught: \cite{devnos_genomics_2016} found that patients are more likely to share their genetic information with Google than insurance companies. The future of life insurance underwriting is expected to become increasingly computational and automated through the usage of AI and the collection of more personalized, individual genetic data \cite{balasubramanian_insurance_2021, filabi_ai-enabled_2021, maier_improving_2020, perumalsamy_ai-driven_2023}. While medical records and demographic factors can be used for mortality analysis, the inclusion of genetic factors can mean that risk prediction can be performed much earlier in an applicant's life without necessarily the same amount of records as an older individual, emphasizing the risk of genetic discrimination \cite{karlsson_linner_genetic_2022} based on the potential to have risk factors (e.g., a genetic risk for high blood pressure versus clinically-measured high blood pressure). Life insurance is not covered by GINA and life insurance companies can access medical records (which may include genetic test results) as part of an application. There is evidence of industry pressure and lobbying to shift the focus of bills to not protect genetic discrimination in life insurance \cite{rothstein_time_2018}. Bills have been introduced in several states that would restrict the use of genetic information in underwriting, but these efforts largely failed: as of 2022, out of thirty-seven proposed bills across all states, three were introduced, eight were signed by the governor, demonstrating the geographical dependency of policy-based protection \cite{vermont_legislature_genetic_2019}.

Life insurance companies may be interested in using PGS to assess an individual's risk for various diseases. In early 2024, U.S.-based life insurance company MassMutual and U.K.-based Genomics PLC announced a partnership, offering free genetic testing to MassMutual's life insurance customers \cite{massmutual_genomics_2024}. However, the press release stressed that MassMutual would not receive individual results and that current premiums/policies would be unaffected. This example particularly highlights the leaky interface between pillars, where the Right of Privacy and Right of Action are called into question. AI insurance underwriting algorithms already suffer from racial biases \cite{lee_ai_2022}. That PGS suffer from the portability problem (itself caused by systematic biases in training datasets) means that these biases could be perpetuated in genetics-informed life insurance underwriting. Additionally, given the relational aspect of genetic data, genetic-based underwriting could affect other biological relatives' applications, harming not only relatives but any others with the same genetic markers that are associated with mortality risks. The vulnerabilities of (1) unsettled science and (2) rapid evolution and usages of genetic data can exacerbate the potential harms of using genetic data as part of underwriting in life insurance.

\section{Recommendations}


\subsection{Recommendation 1: Redefining Genetic Data}

\added{

\textbf{Issue:} Existing policy language does not adequately define genetic data or genetic information.

\textbf{Recommendation:}  Given that genetic data possess unique characteristics distinct from other identifying information, we propose a two-tier definitional framework.

\begin{enumerate}
    \item \textbf{Genetic data }refers to any information relating to an individual's genetic characteristics, including but not limited to DNA or RNA sequences, genomic variants, gene expression profiles, epigenetic markers, or genetic sequences from biological relatives. This data shall always be considered personally identifiable information (PII) as it pertains to unique biological attributes that can potentially be linked to a specific individual, their biological relatives, or identifiable group. For the purposes of this definition, any de-identified, pseudonymized, or anonymized genetic information shall be treated as genetic data, regardless of measures taken to mask individual identities, recognizing that the inherent characteristics of genetic data may enable re-identification through advanced technological or data cross-referencing methods.
    \item \textbf{Genetic information} encompasses any downstream inferences, predictions, prior inferences, or derivations extracted from genetic data, including but not limited to health predispositions, ancestry determinations, trait predictions, and algorithmic outputs based on genetic data analysis.
\end{enumerate}

While genetic information may not be immediately identifiable as traditional PII, we recommend that both genetic data and genetic information be treated as a \emph{distinct protected category for sensitive data} requiring enhanced safeguards and the same level of protections as PII and protected health information (PHI). This approach draws from established legal precedent in biometric privacy law, specifically the distinction between ``biometric identifiers'' and ``biometric information'' established in Illinois's Biometric Information Privacy Act \citep{illinois_bipa_2008}.
}



\subsection{Recommendation 2: Extending Protections for Genetic Discrimination}

\textbf{Issue:} While GINA covers health insurance and employment, it does not protect life, long-term care, and disability insurance, as well as other domains (e.g., housing, education). 

\textbf{Recommendation:} We recommend that additional federal laws should be enacted to extend GINA's coverage beyond employment and health insurance and for GINA to adopt and encompass the definitions of genetic data and genetic information provided in Recommendation 1. The most comprehensive state genetic anti-discrimination law is California's CalGINA, which includes housing, mortgage brokerage, education, and more. However, despite these extensions, CalGINA does not cover life, disability, or long-term care insurance. We recommend that a comprehensive update to GINA should explicitly cover life, disability, long-term care, as well as education and any other opportunities for advancement to be protected against genetic discrimination. Legislation should be written in such a way that prohibits any barrier to opportunity, even those not anticipated at the time of writing. Additionally, it should explicitly bar genetic risk for complex traits known to have significant environmental influences (such as cardiovascular disease) from being considered a preexisting condition.

\subsection{Recommendation 3: A Genetic Data Regulation Framework}

\textbf{Issue:} Current protections do not comprehensively apply to the full genetic data ecosystem which leads to \textit{leaky} protections (e.g., clinical tests being used in life insurance).

\textbf{Recommendation:} We suggest a uniform regulatory framework that applies the same standards of collection, consent, and control to all data collectors in the genetic data ecosystem. In the spirit of the \textit{legitimacy} challenge posed by Viljoen, we build upon specific values inherent in privacy rights to ensure that responsibility is placed on the organizations that hold and analyze data, rather than the individuals from whom it was collected, and the approach to consider data as a democratic medium \cite{solove_limitations_2022, viljoen_relational_2024}. 

\begin{enumerate}
\item \textbf{Entity approval:} Entities—whether they be universities, health care providers, hospital systems, or corporations—that collect or house any type of genetic data (DNA, RNA, or even genetic test results) must have prior regulatory approval. Approval requires a clear commitment to the basic rights of the individuals whose genetic data the entity will collect or own, such as data privacy, security, the right to deletion, etc.
\item \textbf{Test approval:} After entity approval, we propose regulations for the inferential tests, specifically those that return results to either an individual or a medical practitioner for any actionable result. At a minimum, entities would be required to publicly release detailed white papers for each test that detail laboratory procedures, quality control steps, inferential models used to validate the testing process for AI models, presentation of results, and any other necessary details that would allow one to recreate the analysis if given the same data.
\item \textbf{Powers given to the individual}
\begin{itemize}
\item Individuals should be able to request data removal. This would include specifically (1) destruction of the physical sample, (2) deletion of data, (3) removal of any identifiers, and (4) any removed data no longer influence the results of any downstream models.
\item Third-party data transfers should only occur between companies authorized to collect or own genetic data, with recipient companies subject to the same regulations. We recommend providing individuals with the option of blanket opt-out from data transfers and specific opt-in choices for each third party. Individuals must be notified of the transfer and given a reasonable time to opt-out.
\item Research usage is common among DTC companies, academia, and hospitals, where data provided from individual tests can be pooled together as a dataset for research (both internally and via external collaborations). We suggest an opt-out strategy here for individuals, with a specification on whether they consent to their data becoming publicly available in any form (e.g. public datasets, open sourced GWAS weights, trained models, etc).
\item Incidental genetic discoveries can be possible in research use cases where data for a particular test was used for other tests. In these cases, where knowledge of a particular trait can also be harmful and uninformative to the individual, we suggest an opt-in strategy for an individual to blanket choose if they wish to hear any secondary discoveries, or forward any secondary discoveries to their medical practitioner.
\end{itemize}
\item \textbf{Bankruptcy and Acquisition:} As detailed by third-party data transfers, the owning of genetic data would also apply to companies who acquire any genetic data as part of assets through acquisitions or bankruptcy. Thus, all entities involved must already have approval to handle genetic data. In the rare event that there are no entities to handle the data, we suggest a similar protocol structure as nuclear waste, where all physical sample, data, and models are subsequently destroyed and cannot be recovered.
\end{enumerate}

\section{Conclusions and Open Challenges}

In this paper we outline the genetic data ecosystem and today's state of public protections. We identify critical gaps in current regulatory federal and state-level frameworks that prevent the productive, ethical, and safe adoption of genetic data for broad societal applications. To address this, we propose a novel risk assessment framework that offers key public protections and preserve's individuals personal liberties, and finally make three concrete policy recommendations.

While our policy recommendations focus on enhancing individual protections, we also identify three key challenges in regulating genetic data usage that fall outside the scope of our work and that would benefit from further research.

\subsection{The Relational Theory of Privacy}

\subsubsection{Should genetic data submissions require consent from biological relatives?} When an individual does a genetic test they are, indirectly, testing (part of) their relative's genome. What (if any) consent or privacy measures should be taken to protect their biological relatives who may not consent to such a test? Additionally, what should happen upon death? Do children ``own'' their parents' genome (and what ownership would, say, a full sibling have?) and can they make decisions (e.g., to delete the data)?

\subsubsection{Who has ownership of children's DNA?} Do parents have a ``fiduciary duty'' to protect their child's genetic data and protect them from the system vulnerabilities that we have outlined? Part of this fiduciary duty would involve whether a genetic test should be taken in the first place, e.g., the difference between a medically-necessary test and a DTC genetic test for insights into the child's ancestry and genealogical relatives. The relational aspect of genetic data also comes into play; for example, should consent be required from both parents to sequence their child's genome? See \cite{bala_who_2023} for a thorough discussion of child's genome ownership.

\subsection{International Genetic Data Transfers}

Our paper focuses on U.S. regulatory and legislative bodies and U.S.-based companies. Should there be restrictions on genetic data transfer between countries? For example, one hypothetical scenario might involve a political refugee who has medically-necessary genetic testing done within the U.S, but because of leaky data transfer rules, genetic information is transferred to their home country where it might be used to identify family members now at risk for persecution.

\subsection{Collective Rights for Genetic Data}

Privacy scholars acknowledge the necessity of collective rights when building on the relational nature of genetic data. Government-controlled genetic databases present a fundamental challenge to democratic legitimacy, creating power imbalances that distort citizen-state relationships and potentially undermine collective political engagement. Safeguarding individual data-subject rights, while important, does not address the deeper issue: the lack of collective, democratic decision-making regarding whether, how, and for what purposes genetic data is created and used. Key scholars addressing collective rights include King, who demonstrates this phenomenon in DTC genetic testing \cite{king_becoming_2019}; Solow-Neiderman, who details how individual genetic submissions compromise collective privacy \cite{solow-niederman_information_2021}; Costello, who raises issues of group consent and responsibility \cite{costello_genetic_2022}; and Viljoen, who argues for democratizing data for the public collective \cite{viljoen_relational_2024}.


\added{While robust theoretical arguments support collective rights frameworks, their path to implementation is less clear. Our analysis bridges existing policy approaches with emerging theoretical frameworks, providing a scaffold for comprehensive genetic data governance. In an increasingly data-driven society, there is an urgent need to address major gaps in genetic data governance, which informs our process to develop a risk assessment framework and outline three concrete policy recommendations. This work represents not an endpoint but a call to action for policymakers, researchers, and ethicists to collaborate on transforming theoretical insights into practicable governance solutions that safeguard genetic privacy for present and future generations.}   



\newpage

\bibliography{genetic_privacy} 

\end{document}